\documentclass[conference]{IEEEtran}
\IEEEoverridecommandlockouts

\usepackage{cite}
\usepackage{amsmath,amssymb,amsfonts}
\usepackage{algorithmic}
\usepackage{graphicx}
\usepackage{textcomp}
\usepackage{xcolor}
\def\BibTeX{{\rm B\kern-.05em{\sc i\kern-.025em b}\kern-.08em
    T\kern-.1667em\lower.7ex\hbox{E}\kern-.125emX}}
\begin{document}

\vspace{1em}
\title{A Proxy-Based Method for Mapping Discrete Emotions onto VAD model\\

\thanks{This work was partially financially supported by Gdańsk University of Technology under the grant DEC-35/1/2024/IDUB/IV.2a/Eu within the Europium -- ‘Excellence Initiative - Research University’ program.}
}

\author{\IEEEauthorblockN{1\textsuperscript{st} Michal R. Wrobel}
\IEEEauthorblockA{\textit{Gdansk University of Technology} \\
\textit{ Faculty of Electronics, Telecommunications 
and Informatics; Digital Technologies Center} \\
Gdańsk, Poland \\
0000-0002-1117-903X}
}

\maketitle
\textbf{ Preprint Notice}
This manuscript is a preprint of an article accepted for presentation at \textit{The 24th IEEE/WIC International Conference on Web Intelligence and Intelligent Agent Technology (WI-IAT 2025}), London, UK. The final published version may differ as a result of the publisher’s production process.\\[2mm]

\begin{abstract}
Mapping discrete and dimensional models of emotion remains a persistent challenge in affective science and computing. This incompatibility hinders the combination of valuable data sets, creating a significant bottleneck for training robust machine learning models. To bridge this gap, this paper presents a novel, human-centric, proxy-based approach that transcends purely computational or direct mapping techniques. Implemented through a web-based survey, the method utilizes simple, user-generated geometric animations as intermediary artifacts to establish a correspondence between discrete emotion labels and the continuous valence-arousal-dominance (VAD) space. The approach involves a two-phase process: first, each participant creates an animation to represent a given emotion label (encoding); then, they immediately assess their own creation on the three VAD dimensions. The method was empirically validated and refined through two iterative user studies. The results confirmed the method's robustness. Combining the data from both studies generated a final, comprehensive mapping between discrete and dimensional models. 
\end{abstract}


\section{Introduction}
Accurately representing human emotions is a fundamental challenge in fields ranging from psychology to affective computing and human-computer interaction. Two paradigms have historically dominated this landscape: the discrete model, which uses intuitive, categorical labels such as ``joy'', ``sadness'',and ``anger'' to describe emotions, and dimensional models, which characterize affect as a point within a continuous, multidimensional space. One of the most prominent dimensional models is the Valence-Arousal-Dominance (VAD) model, which captures the pleasantness, intensity, and sense of control associated with an emotional state. Discrete labels offer ecological validity and ease of use for human reporting. However, the VAD model provides a nuanced, quantitative framework that is highly advantageous for computational analysis and processing. However, bridging the conceptual gap between these two representations remains a significant issue. A robust method for translating intuitive emotional labels into meaningful dimensional coordinates is vital for developing technologies that can recognize and respond to the entire range of human emotions.

This need for a translational framework is particularly acute in the context of machine learning and data-driven AI. Although the field of affective computing has many datasets that can be used to train emotion recognition models, these resources are fragmented by their annotation schemes. Some datasets are labeled using discrete categories, while others are annotated with continuous VAD values. This incompatibility prevents researchers from merging these resources to create larger, more comprehensive, and diverse training corpora, which could help build robust, generalizable models. Consequently, developing sophisticated emotion-aware systems is often hindered by the limitations of smaller, siloed datasets. A robust, psychologically grounded mapping method would be a valuable tool for harmonizing these data resources, unlocking the potential to advance automated emotion recognition significantly.

To address these challenges, this paper introduces a novel, human-centric method that goes beyond purely computational or direct mapping techniques. This method relies on a survey to collect and aggregate data from multiple participants, forming the basis of the final mapping.  We propose using a simple, user-generated geometric animation as an intermediary proxy to establish a correspondence between discrete labels and the VAD space.  The method involves a two-phase process that each participant completes. First, participants translate a given emotion label into a dynamic visual representation by selecting parameters such as object shape, color, and animation type. Then, they assess their newly created animation on the continuous dimensions of valence, arousal, and dominance. The animation serves as a nonverbal, personalized artifact that represents an individual's interpretation of an emotion. Such mapping is grounded in a person's creative expression and analytical judgment.

The remainder of this paper is organized as follows. First, related work on established emotion models and the challenges associated with mapping between them is reviewed to situate the present contribution. Next, in the Section~\ref{sec:method}, the novel proxy-based method for mapping emotions through user-generated animation is introduced. In Section~\ref{sec:validation}, two user studies conducted in an iterative process to validate and refine the method are presented: the first study establishes its fundamental feasibility with a constrained interface, while the second confirms its robustness using a design that offers greater expressive freedom. Next, validated mapping of a set of common emotion labels onto the VAD space is presented. Finally, the results are discussed in Section~\ref{sec:discuss} and the work is concluded in Section~\ref{sec:conclude}.

\section{Background and Related Work}\label{sec:related}

Historically, the computational representation of human emotion, a key component of affective computing, has been approached through two dominant paradigms: discrete and dimensional models \cite{kolakowska2015modeling}. The first of these is the discrete model, which posits that emotions can be classified into a finite set of distinct, universally recognized categories, such as the six basic emotions proposed by Ekman \cite{ekman1971constants}. These categories are usually described with intuitive ``joy'', ``sadness'', and ``anger'', which makes them easy for humans to report and annotate. In contrast, dimensional models characterize affect as a point within a continuous, multidimensional space rather than as separate states. One of the most prominent is the valence-arousal-dominance (VAD) model, \cite{Russell1977}, which quantifies an emotional experience along three axes: valence (pleasantness), arousal (activation), and dominance (control). This model is particularly advantageous for the computational processing of emotions in modern affect-aware systems.

Rather than viewing these two paradigms as mutually exclusive, a growing body of research suggests they provide complementary perspectives that are essential for comprehensively understanding affect. Discrete emotions provide functional specificity, and dimensional models offer insight into the underlying structure and similarities between different affective states. Serious consideration of both approaches has been shown to be highly beneficial. For instance, it helps reveal how affective dimensions can vary within a single discrete emotion or how different discrete emotions can share underlying motivational characteristics~\cite{Harmon-Jones2017Sep}. This synergy underscores the value of integrating insights from both models, as neither paradigm on its own may be sufficient to capture the full complexity of human emotion. Therefore, the ability to translate between these representations is a crucial step toward a more holistic and functional model of affect, not merely a technical convenience.

However, the coexistence of these two prevalent modeling paradigms gives rise to significant practical and theoretical challenges. The structural differences between categorical labels and dimensional coordinates create a fundamental incompatibility that hinders the interoperability of systems, the comparability of research findings, and the fusion of affect-related data. Often referred to as the ``mapping problem,'' this issue has been identified as a key obstacle in the development of affect-aware applications~\cite{kolakowska2015modeling}. The challenge becomes especially acute in areas such as multimodal emotion recognition, where data streams from different sources, each potentially analyzed using a different emotion model, must be integrated to form a coherent assessment of a user's state~\cite{landowska2018towards}.

This representational divide has especially severe consequences in the data-driven field of affective computing. Developing robust, generalizable machine learning models for emotion recognition depends heavily on large-scale, diverse training corpora. However, available data resources such as annotated lexicons, image databases, and speech corpora are often fragmented by their annotation schemes. The growing number of incompatible resources decreases their reusability and hinders direct comparisons of systems built upon them~\cite{buechel2018representation}. Researchers are therefore often unable to merge valuable datasets to create larger, more varied training sets needed to advance the state of the art. Furthermore, the absence of a common representational ground complicates the crucial task of benchmarking. For example, evaluating a system that outputs VAD scores against a ground truth annotated with basic emotions presents a significant methodological challenge ~\cite{buechel2016emotion}.

In response to this challenge, the predominant approach in the literature has been to derive mappings computationally by identifying statistical relationships within large, annotated corpora. WIn the field of natural language processing, this often involves using affect-annotated lexicons to train regression models that predict VAD scores from discrete emotion labels, or vice versa~\cite{landowska2018towards, buechel2018representation, calvo2013emotions}.  Similar data-driven techniques have been successfully applied to emotional speech. For example, Trnka et al. applied machine learning models with X-vector representations that were trained to predict dimensional values directly from the acoustic features of utterances labeled with categorical emotions~\cite{Trnka2021Nov}. This methodology can be applied to visual data as well. For example, Horvat et al.'s study demonstrated that it is possible to statistically infer dimensional ratings for affective images based on their discrete emotion annotations~\cite{Horvat2022Aug}. While these computational methods are demonstrably effective, the resulting mappings are constrained to specific modalities (e.g., text, speech, or images) and the statistical properties of the datasets utilized for training, which may limit their generalizability.

A different, more direct approach to the mapping problem bypasses modality-specific datasets and instead relies on human psychological judgment. This method involves asking participants to translate emotion labels into dimensional coordinates based on their own subjective understanding. In a notable example of this approach, Hoffmann et al.~\cite{hoffmann2012mapping} instructed participants to place the labels of emotions directly into the three-dimensional VAD space using a simple visual tool. Their findings revealed that, while a high degree of inter-subject consistency was observed for the placement of emotions along the pleasure (valence) dimension, the ratings for arousal and dominance showed considerably greater variance. These results suggest that, although the valence of an emotion label is a well-understood and consistently mapped concept, directly translating a word into specific levels of arousal and dominance is a more difficult, subjective, and less reliable task.

A review of the state-of-the-art reveals a critical trade-off between the two primary mapping strategies. On one hand, computational methods produce mappings that are data-driven and modality-specific, but they are fundamentally indirect, reflecting statistical correlations within a given dataset rather than a direct translation of human affective understanding. On the other hand, the direct psychological judgment approach is more fundamental and modality-independent, but it appears to place a significant cognitive load on participants. The difficulty of translating an abstract emotion concept directly into precise dimensional coordinates is highlighted by the high variance in arousal and dominance ratings found in previous work \cite{hoffmann2012mapping}. This suggests a clear gap in the existing methodologies: a need for an approach that is psychologically grounded like the direct judgment method, but which mitigates the cognitive difficulty that leads to unreliable assessments, particularly for the more elusive dimensions of arousal and dominance.

\section{The Proxy-Based Mapping Method}\label{sec:method}

The proposed method is designed to create a psychologically grounded mapping between discrete emotion labels and the dimensional VAD space. The core of this approach is the use of an intermediary artifact, or ``proxy'', to facilitate the translation. This proxy takes the form of a simple, user-generated geometric animation. The rationale for this choice is rooted in the established human capacity to attribute affective meaning to abstract visual stimuli, including shapes, colors, and patterns of motion. Rather than requiring participants to perform the often difficult cognitive task of translating a word into a set of abstract dimensional ratings, this method externalizes emotions into dynamic, nonverbal representations that can be assessed.

The mapping process is divided into two distinct, sequential phases, each of which is completed by the same participant for each emotion label.

\begin{enumerate}
    \item \textbf{Phase 1: Emotion Label to Animation (Encoding)}: A participant is presented with a discrete emotion label (e.g., ``joy''). Their task is to create a simple animation that, in their view, best represents this emotion. They do this using a purpose-built tool that allows them to select an object's shape, color, size, and type of animation. The output of this phase is an animation that serves as the proxy for the initial emotion label.
    \item \textbf{Phase 2: Animation to VAD (Assessment)}: After creating all animations, participants are tasked with assessing their own creations using the three dimensions of the VAD model. They provide scores for valence (pleasantness/unpleasantness), arousal (excitement/calmness), and dominance (control/submissiveness) using standard rating scales.
\end{enumerate}

By completing this two-phase sequence, a direct link is established between the initial discrete label from Phase 1 and the final VAD coordinates from Phase 2. The user-generated animation is the crucial bridge that ensures the final mapping is grounded in the participant's integrated process of creative expression and subsequent perceptual judgment rather than based on abstract semantic association. For each emotion label, this process yields a complete data record consisting of the chosen animation parameters and the corresponding VAD scores.

\section{User Studies: An Iterative Approach to Validation and Mapping}\label{sec:validation}
To empirically validate the proposed animation proxy method and generate a reliable emotion map, we conducted two user studies in an iterative sequence. The first study assessed the method's fundamental feasibility and reliability using an interface with a limited color palette and other constrained expressive options. Based on the insights from the initial validation, the second study aimed to replicate the findings while refining the methodology by providing participants with greater expressive freedom via an unconstrained color selection tool. This section details the methodology and results of each study, demonstrating the iterative process that led to the final validated mapping of emotion labels to the VAD space.

\begin{figure}[htbp]
\centerline{\includegraphics[width=0.49\textwidth]{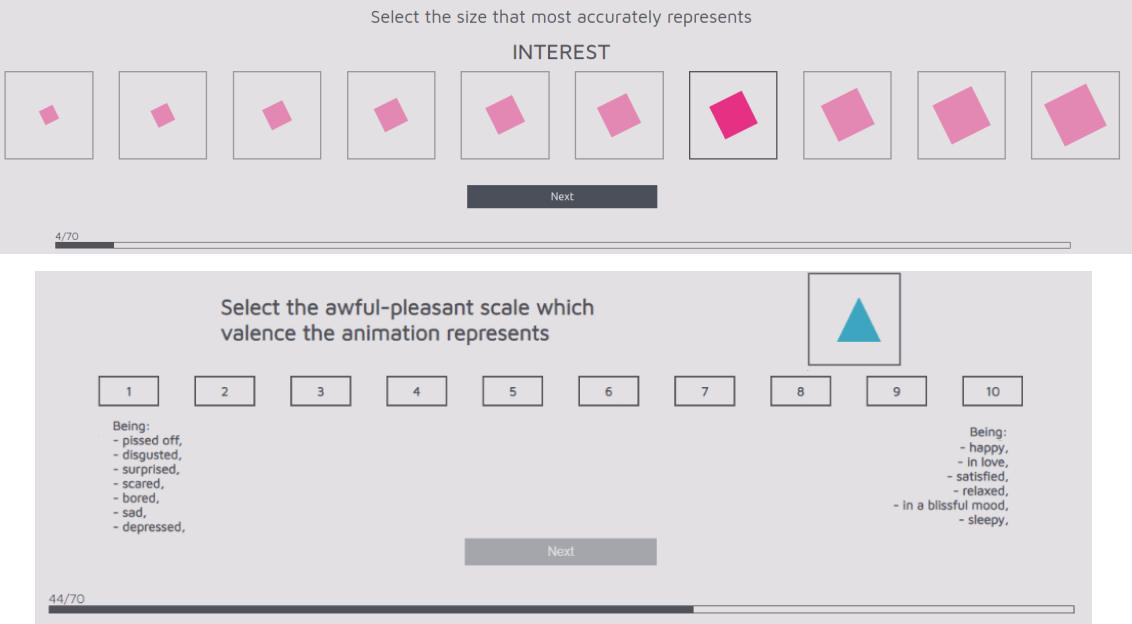}}
\caption{Sample screens from Study 1 web-based questionnaire}
\label{fig:app1}
\end{figure}

\subsection{Study 1: Initial Validation}\label{study1}
The primary aim of the first study was to establish the feasibility and reliability of the animation proxy method. Specifically, the study was designed to determine if the method could successfully distinguish between different emotions using a limited set of expressive tools. The study tested the hypothesis that animations generated for distinct emotion labels would produce significantly different and coherent patterns of scores on the valence, arousal, and dominance dimensions. The initial validation was conducted using a constrained version of the user interface featuring a preselected palette of ten colors to serve as a proof of concept for the core methodology. For the purpose of this study, Izard's model~\cite{izard1977differential} was adopted, which includes the emotions ``interest'', ``joy'', ``surprise'', ``sadness'', ``anger'', ``disgust'', ``contempt'', ``fear'', ``shyness'', and ``guilt''. This model was chosen due to its manageable number of labels and inclusion of both basic and socially oriented emotions, enabling systematic yet diverse emotional representation in the animations.

Figure~\ref{fig:app1} shows a web-based application that was developed to allow participants to complete the two-phase mapping process. The intuitive interface guides users sequentially through the animation creation and VAD assessment stages. When presented with an emotion label, participants entered a multi-page workflow to create their animation. Each core parameter was configured on a separate page to provide a step-by-step process. The sequence was as follows:

\begin{itemize}
    \item \textbf{Animated Object}: Participants select an animated geometric shape from a set of ten options: a blinking rectangle, a squeezing rectangle, a bouncing square, a rotating square, a rotating and enlarging square, a jumping triangle, a bouncing triangle, an enlarging circle, a jumping circle, and a rotating circle.
    \item \textbf{Object Size}: Participants adjust the size of the object by selecting one of the ten available options.
    \item \textbf{Animation Speed}: Participants adjust the speed of the animation by pointing to one of the ten fields containing an animated object.
    \item \textbf{Object Color}: Participants select the color of an object. For this study, the color selection was limited to a palette of ten preselected colors based on Plutchik's theory of emotions\cite{Plutchik1980Jan}: red, pink, violet, navy, steel, orange, khaki, green, gray, and blue.
\end{itemize}

After finalizing the animation, the participants automatically proceeded to the assessment phase. The interface displayed their newly created animation at the top of the screen. On the next three screens, they assessed the animation in terms of valence, arousal, and dominance. Each dimension was rated using a 10-point scale. The scales were anchored with descriptive labels and corresponding numbers:

\begin{itemize}
    \item Valence ranged from 1: ``very unpleasant'' to 10: ``very pleasant''.
    \item Arousal ranged from 1: ``very calm'' to 10: ``very excited''.
    \item Dominance ranged from 1: ``without any control'' to 10: ``in complete control''.
\end{itemize}

A total of 419 Polish nationals successfully completed the study. The sample was composed of 150 participants who identified as female (35.8\%), 261 as male (62.3\%), and 5 as other (1.2\%); three participants (0.7\%) did not specify their gender. The age distribution of the sample was not even, skewing towards younger adults, with the vast majority of respondents aged between 18 and 32 years.

The VAD scores for each emotion label, summarized in Table~\ref{tab:study1-results}, reveal distinct and psychologically coherent affective profiles. ``Joy'' emerged as the quintessential positive, high-energy state, with the highest valence score (M=6.99) and very high scores for arousal (M=7.61) and dominance (M=6.54). In stark contrast, ``anger'' was characterized by its combination of low valence (M=3.58) with the highest levels of both arousal (M=8.07) and dominance (M=7.84), reflecting a state of energized control. On the other end of the spectrum, ``sadness'' was distinguished by low scores across all three dimensions, particularly in arousal (M=3.01) and dominance (M=3.51). The standard deviations were relatively consistent across all emotions and dimensions, generally falling within a range of 2.1 to 2.5. This suggests that, although the central tendencies for each emotion are clearly differentiated, there is a substantial and comparable degree of individual variation in how participants express and perceive these affective states. Together, these well-differentiated profiles strongly support the primary hypothesis of Study 1: the animation proxy method is a feasible and effective tool for discriminating between discrete emotions.

\begin{table*}[ht]
\centering
\caption{Mean and Standard Deviation (SD) of VAD scores for each emotion reported in Study 1.}
\label{tab:study1-results}
\begin{tabular}{l cc cc cc}
\hline
& \multicolumn{2}{c}{\textbf{Valence}} & \multicolumn{2}{c}{\textbf{Arousal}} & \multicolumn{2}{c}{\textbf{Dominance}} \\
\hline
\textbf{Emotion} & \textbf{Mean} & \textbf{Standard Deviation (SD)} & \textbf{Mean} & \textbf{Standard Deviation (SD)} & \textbf{Mean} & \textbf{Standard Deviation (SD)} \\
\hline
anger & 3.584 & 2.382 & 8.071 & 2.141 & 7.843 & 2.147 \\
contempt & 4.819 & 2.246 & 4.290 & 2.213 & 4.876 & 2.366 \\
disgust & 4.221 & 2.123 & 5.428 & 2.287 & 4.606 & 2.216 \\
fear & 3.964 & 2.300 & 6.466 & 2.461 & 4.881 & 2.421 \\
guilt & 5.375 & 2.097 & 4.327 & 2.214 & 4.675 & 2.179 \\
interest & 6.187 & 2.186 & 6.170 & 2.451 & 5.478 & 2.317 \\
joy & 6.993 & 2.409 & 7.613 & 2.172 & 6.542 & 2.263 \\
sadness & 4.024 & 2.326 & 3.007 & 2.008 & 3.508 & 2.377 \\
shyness & 5.831 & 2.281 & 3.565 & 2.302 & 3.375 & 2.183 \\
surprise & 4.230 & 2.125 & 6.150 & 2.521 & 5.895 & 2.466 \\
\hline
\end{tabular}
\end{table*}

\subsection{Study 2: Replication and Refinement}\label{study2}
\begin{figure}[htbp]
\centerline{\includegraphics[width=0.45\textwidth]{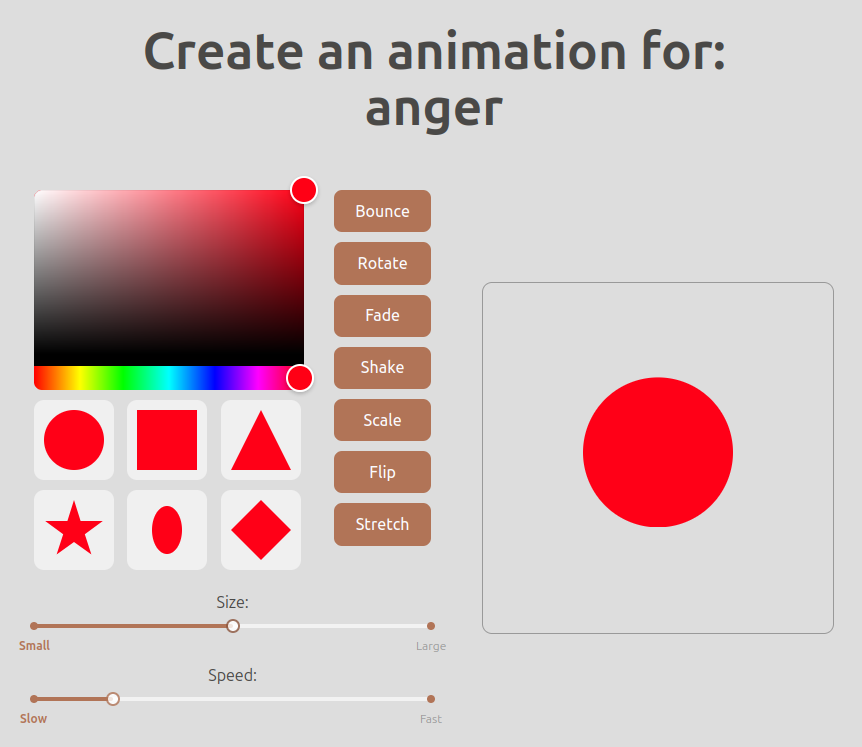}}
\caption{Study 2 web-based questionnaire}
\label{fig:app2}
\end{figure}

Building upon the foundational results of Study 1, the second study was designed with two primary objectives: replication and refinement. The first objective was to confirm the robustness of the animation proxy method by determining if the core finding that the method can reliably discriminate between distinguished emotions could be reproduced with a new sample of participants. The second objective was to investigate the impact of enhancing users' expressive freedom on the mapping process. Three significant modifications were implemented in the web-based tool to this end: first, the constrained 10-color palette was replaced with a full RGB color picker; second, the sequential, multi-page creation workflow was redesigned into a single-page interface where all parameters could be adjusted simultaneously; and third, the set of creative options was expanded to allow participants to construct an animation by combining one of six geometric shapes (circle, star, rectangle, rhombus, triangle, and oval) with one of six animation types (bounce, rotate, fade, shake, flip, and stretch). The user interface of the tool is presented in Figure~\ref{fig:app2}. The aim was to test whether this more fluid and expressive interface would not only replicate the initial findings but also lead to more nuanced emotional representations, thereby producing a more definitive and robust final emotion map. To ensure a direct comparison for this replication and analysis, the same set of emotional labels was used.

The participant pool for the second study consisted of 105 university students. The sample was comprised of 51 students from Poland and 54 from Turkey. The age of all participants was within a narrow range of 19 to 23 years, creating a homogenous sample in terms of age and educational background. As the primary aim of this study was to test the refined methodology rather than conduct a cross-cultural comparison, and preliminary analyses indicated no significant main effect of nationality on the VAD ratings, the data from both groups were pooled for all subsequent analyses. This approach allowed for a more robust statistical evaluation of the refined method's performance on a single, larger dataset.

The results from Study 2, presented in Table~\ref{tab:study2-results}, confirm the overall findings of the initial study while also revealing the significant impact of the methodological refinements. The fundamental structure of the emotion map was replicated, with ``anger'' remaining a high-arousal and high-dominance state and ``sadness'' a low-arousal and low-dominance one. However, notable shifts occurred, most prominently along the valence dimension. For instance, ``joy'' achieved a much higher valence score (M=8.81 compared to 6.99 in Study 1), while negative emotions such as ``anger'' (M=2.59 vs. 3.58) and ``fear'' (M=2.21 vs. 3.96) were rated as significantly more unpleasant. The most striking change was observed for ``surprise,'' which shifted from a negative-leaning valence in Study 1 (M=4.23) to a clearly positive one in Study 2 (M=7.05). These systematic shifts are strongly associated with the replacement of the constrained color palette with a full RGB picker. This indicates that allowing participants to select more subtle, muted, or saturated colors enabled a more nuanced and accurate mapping of the valence dimension, an expressive capability that was limited in the initial study.

A second key finding relates to the variability of the data. In contrast to Study 1, the standard deviations in Study 2 were generally higher across most emotions and dimensions. This suggests that the increased expressive freedom afforded by the refined interface, particularly the unconstrained color choice and single-page design, allowed for a wider and more diverse range of emotional expressions. While the core concepts of the emotions were preserved, participants explored a larger portion of the affective space in their creations. This greater variance is not a sign of reduced reliability but rather an indicator that the refined tool was more successful at capturing the rich and varied subjective interpretations of each emotion. In conclusion, these results not only validate the robustness of the proxy method but also demonstrate that enhancing the user's expressive toolset is critical for capturing the fine-grained details and the full range of individual differences in the affective space.

\begin{table*}[ht]
\centering
\caption{Mean and Standard Deviation (SD) of VAD scores for each emotion reported in Study 2.}
\label{tab:study2-results}
\begin{tabular}{l cc cc cc}
\hline
& \multicolumn{2}{c}{\textbf{Valence}} & \multicolumn{2}{c}{\textbf{Arousal}} & \multicolumn{2}{c}{\textbf{Dominance}} \\
\hline
\textbf{Emotion} & \textbf{Mean} & \textbf{Standard Deviation (SD)} & \textbf{Mean} & \textbf{Standard Deviation (SD)} & \textbf{Mean} & \textbf{Standard Deviation (SD)} \\
\hline
anger & 2.585 & 2.346 & 8.057 & 2.429 & 8.170 & 2.606 \\
contempt & 3.981 & 2.828 & 5.010 & 3.021 & 5.733 & 2.916 \\
disgust & 2.952 & 2.105 & 5.010 & 2.684 & 4.419 & 2.931 \\
fear & 2.208 & 1.766 & 6.802 & 3.016 & 4.972 & 3.220 \\
guilt & 2.914 & 2.497 & 4.676 & 3.136 & 3.971 & 3.254 \\
interest & 7.362 & 2.206 & 6.095 & 2.677 & 5.286 & 2.243 \\
joy & 8.811 & 1.691 & 7.340 & 3.020 & 6.274 & 2.830 \\
sadness & 2.962 & 2.305 & 3.113 & 2.451 & 3.943 & 3.033 \\
shyness & 3.886 & 2.509 & 4.086 & 2.968 & 3.238 & 2.702 \\
surprise & 7.047 & 2.758 & 6.972 & 2.803 & 5.840 & 2.788 \\
\hline
\end{tabular}
\end{table*}

\subsection{Final Emotion Mapping }
After successfully validating the core methodology in Study 1 and demonstrating its robustness in Study 2, a final analysis was conducted to create a comprehensive emotion map. To produce the most statistically powerful and reliable estimates of the VAD coordinates for each emotion label, the datasets from both studies were combined. The consistent patterns observed across both experiments justify this approach. While Study 2's refined interface allowed for greater expressive nuance, the fundamental relationships between specific emotion labels and their corresponding locations within the VAD space remained stable. Pooling the data increased the sample size, providing a more robust foundation for the final map and minimizing the impact of individual outliers. This yielded a more stable and generalizable result. The comprehensive mapping derived from the combined dataset is presented below and represents this work's primary contribution.

\begin{figure*}[htbp]
\centerline{\includegraphics[width=0.58\textwidth]{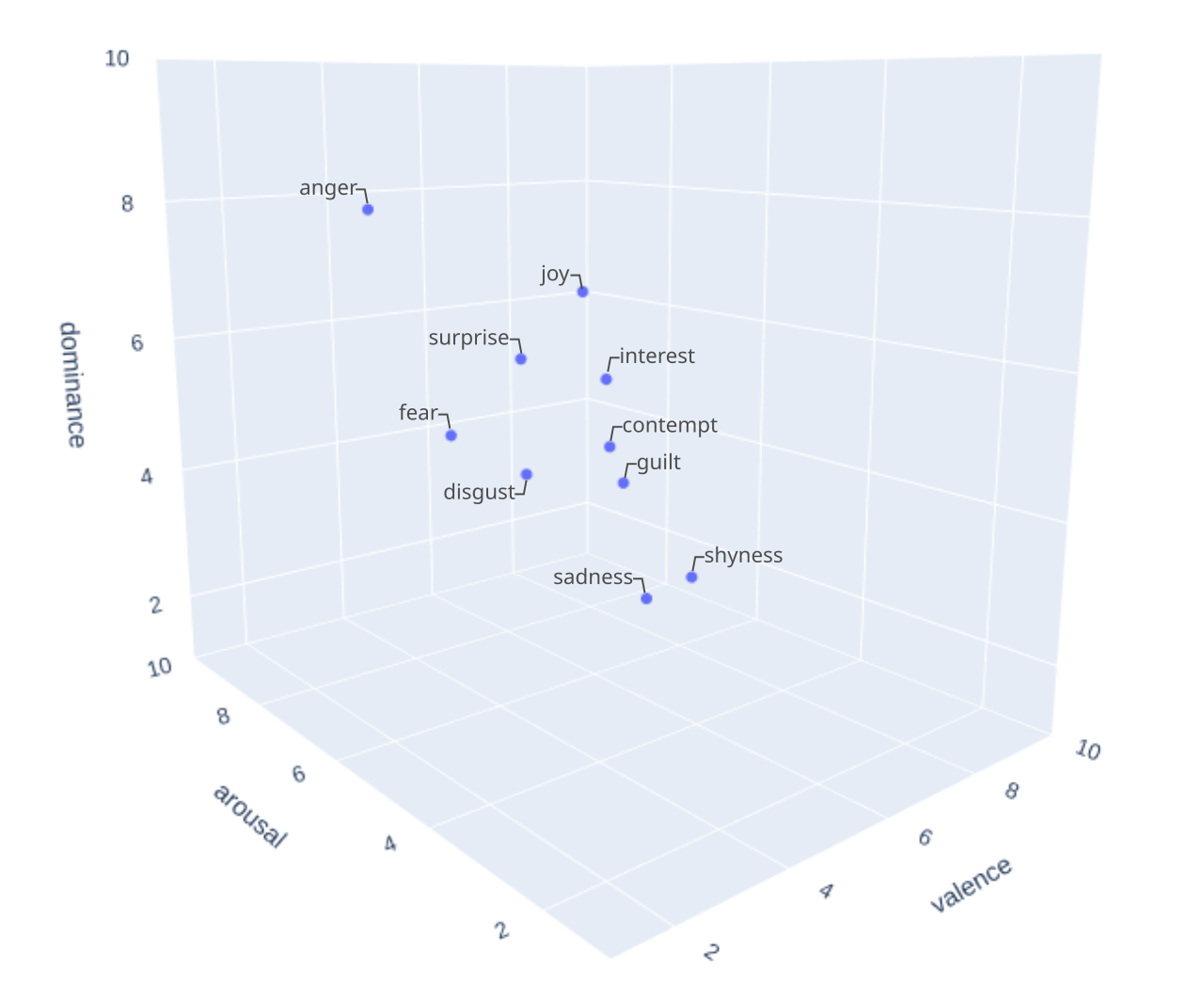}}
\caption{Visualization of mapping between discrete emotions and VAD in 3D space}
\label{fig:mapping}
\end{figure*}

The final VAD coordinates for the ten emotion labels resulting from the combined analysis are visualized in Table~\ref{tab:zscore-analysis} and Figure~\ref{fig:mapping}. The mapping reveals a clear and psychologically coherent structure within the VAD space. As anticipated, ``joy'' is positioned as the quintessential positive state, characterized by high valence (V=7.36, for z-score threshold 3), high arousal (A=7.56), and high dominance (D=6.49). Conversely, a cluster of negative emotions occupies the low-valence region, but with distinct arousal and dominance profiles. ``Anger'' is particularly notable, defined by extremely high arousal (A=8.10) and the highest dominance score (D=8.00), reflecting a state of energized control. ``Sadness'' occupies the opposite corner of the affective space, distinguished by low arousal (A=2.99) and low dominance (D=3.57). ``Shyness'' exhibits a similar low-arousal and low-dominance profile but is positioned closer to neutral valence. Other key emotions like ``fear'' and ``surprise'' are characterized by high arousal but fall into the mid-to-low range for both valence and dominance. This is consistent with their conceptualization as alerting and uncertain states. Overall, the spatial distribution of these emotion labels aligns well with established affective models, demonstrating the method's capacity to produce a nuanced and valid mapping.

To assess the robustness of the final mapping and ensure the results were not unduly influenced by outliers, a sensitivity analysis was performed. The combined dataset was progressively filtered using three different z-score thresholds (3, 2, and 1) to remove extreme or anomalous data points. The resulting mean VAD coordinates for each emotion label at each filtering level are presented in Table~\ref{tab:zscore-analysis}. The analysis reveals a high degree of stability in the mapping. While stricter filtering (a lower z-score threshold) slightly accentuates the affective profiles of certain emotions—for instance, making ``anger'' more extreme in its high-arousal, high-dominance position—the core structure of the map and the relative positioning of the emotions remain remarkably consistent across all conditions. ``Joy'' consistently occupies the high-valence, high-arousal quadrant, while ``sadness'' remains in the low-valence, low-arousal, and low-dominance space. This stability demonstrates that the generated emotion map is a robust representation of the underlying data, not merely an artifact of a few influential outliers, thereby validating the integrity of the final mapping.

\begin{table*}[ht]
\centering
\caption{Mean and Standard Deviation (SD) of VAD scores for each emotion after data cleaning with different z-score thresholds. The number of remaining data points (n) is shown for each threshold.}
\label{tab:zscore-analysis}
\begin{tabular}{l ccc ccc}
\hline
& \multicolumn{2}{c}{\textbf{Valence}} & \multicolumn{2}{c}{\textbf{Arousal}} & \multicolumn{2}{c}{\textbf{Dominance}} \\
\hline
\textbf{Emotion} & \textbf{Mean} & \textbf{Standard Deviation (SD)} & \textbf{Mean} & \textbf{Standard Deviation (SD)} & \textbf{Mean} & \textbf{Standard Deviation (SD)} \\
\\
\multicolumn{7}{l}{\textbf{Z-score threshold = 3 (n=5254)}} \\
\hline
anger & 3.393 & 2.395 & 8.100 & 2.160 & 7.996 & 2.120 \\
contempt & 4.652 & 2.394 & 4.433 & 2.410 & 5.048 & 2.506 \\
disgust & 3.968 & 2.177 & 5.344 & 2.375 & 4.568 & 2.374 \\
fear & 3.611 &  2.312 & 6.533 & 2.583 & 4.899 & 2.599 \\
guilt & 4.882 & 2.393 & 4.397 & 2.427 & 4.534 & 2.445 \\
interest & 6.420 & 2.238 & 6.155 & 2.495 & 5.439 & 2.302 \\
joy & 7.359 & 2.395 & 7.558 & 2.367 & 6.488 & 2.387 \\
sadness & 3.794 & 2.348 & 2.989 & 2.041 & 3.574 & 2.512 \\
shyness & 5.443 & 2.453 & 3.669 & 2.455 & 3.348 & 2.293 \\
surprise & 4.797 & 2.530 & 6.315 & 2.520 & 5.884 & 2.531 \\
\\
\multicolumn{7}{l}{\textbf{Z-score threshold = 2 (n=4863)}} \\
\hline
anger & 3.052 & 2.102 & 8.522 & 1.646 & 8.348 & 1.681 \\
contempt & 4.647 &  2.288 & 4.293 & 2.242 & 4.869 & 2.397 \\
disgust & 3.852 & 2.013 & 5.299 & 2.362 & 4.425 & 2.244 \\
fear & 3.475 & 2.130 & 6.612 & 2.502 & 4.871 & 2.602 \\
guilt & 4.796 & 2.196 & 4.236 & 2.269 & 4.413 & 2.313 \\
interest & 6.434 & 2.211 & 6.237 & 2.434 & 5.452 & 2.287 \\
joy & 7.529 & 2.192 & 7.850 & 2.014 & 6.619 & 2.242 \\
sadness & 3.576 & 2.127 & 2.784 & 1.786 & 3.210 & 2.205 \\
shyness & 5.410 & 2.413 & 3.317 & 2.100 & 3.004 & 1.894 \\
surprise & 4.464 & 2.213 & 6.226 & 2.500 & 5.855 & 2.489 \\
\\
\multicolumn{7}{l}{\textbf{Z-score threshold = 1 (n=1528)}} \\
\hline
anger & 2.271 & 1.363 & 9.006 & 1.199 & 8.982 & 1.054 \\
contempt & 4.870 & 1.207 & 4.565 & 1.025 & 5.428 & 1.093 \\
disgust & 3.943 & 1.274 & 5.019 & 1.430 & 4.302 & 1.078 \\
fear & 3.706 & 1.083 & 5.945 & 1.239 & 4.725 & 1.053 \\
guilt & 4.957 & 1.243 & 4.428 & 1.126 & 4.543 & 1.026 \\
interest & 6.547 & 1.105 & 6.173 & 1.335 & 5.396 & 0.983 \\
joy & 7.427 & 1.228 & 7.359 & 1.197 & 6.557 & 1.131 \\
sadness & 3.519 & 1.309 & 3.074 & 1.055 & 3.119 & 1.140 \\
shyness & 5.087 & 1.333 & 3.559 & 1.117 & 3.315 & 1.132 \\
surprise & 4.961 & 1.336 & 6.055 & 1.323 & 5.528 & 1.068 \\
\hline
\end{tabular}
\end{table*}

\section{Discussion}\label{sec:discuss}
The present work introduced and validated a novel proxy-based method for mapping discrete emotion labels to the continuous VAD space. Through two iterative user studies, the research demonstrated that user-generated geometric animations can reliably and psychologically ground this complex translation. The first study showed the fundamental feasibility of the approach by demonstrating that the method could effectively discriminate between core emotions despite the constrained user interface. The second study confirmed these findings' robustness and emphasized the importance of giving users more expressive freedom, resulting in a final, comprehensive emotion map.

The final mapping, which combines the data from both studies, reveals a spatial organization of emotions that is highly consistent with established affective theories. The location of ``joy'' in the high-valence, high-arousal quadrant and ``sadness'' in the low-valence, low-arousal, low-dominance quadrant aligns perfectly with their conceptual understanding and serves as a strong validation of the method's external validity. More importantly, the method demonstrates a capacity for nuance. For instance, ``anger'' and ``fear'' are both high-arousal, negative-valence emotions, but they are clearly distinguished by dominance; ``anger'' is characterized by a very high sense of control (D=8.00), whereas ``fear'' is associated with a lower sense of control (D=4.90). This distinction is critical and effectively captured by the animation proxy, where users likely expressed ``anger'' with strong, determined motion and ``fear'' with more chaotic or submissive motion. The stability of these relative positions, as confirmed by the z-score sensitivity analysis, underscores that the resulting map is a robust and meaningful representation, not an artifact of outliers.

The iterative design of the two studies revealed a key insight about the method itself: the importance of expressive freedom. The transition from the constrained 10-color palette of Study 1 to the full RGB picker of Study 2 was more than just a technical upgrade. It enabled participants to convey the specific emotional tone associated with an emotion more precisely. Similarly, shifting from a sequential, multi-page creation process to an integrated, single-page interface likely facilitated a more holistic and fluid creative process. While the core findings were successfully replicated, suggesting that the method is robust even in its simplest form, the refinement process demonstrates that the quality and nuance of the mapping are directly related to the quality of the expressive tools provided to users. This has significant implications for designing future studies in human-computer interaction and affective computing that rely on user-generated content.

The primary strength of the proposed method is its human-centric nature. Unlike computational or semantic approaches, which map words to coordinates, this method establishes the mapping based on a participant's creative expression and perceptual judgment processes. The animation is a nonverbal artifact that externalizes an individual’s internal representation of an emotion. This has two key advantages. First, it captures a more vivid, embodied understanding of an emotion that transcends simple dictionary definitions. Second, the two-phase process, in which a participant first creates and then immediately assesses their own work, establishes an internally consistent feedback loop. VAD ratings are not an abstract judgment of a word but rather a concrete assessment of a stimulus that participants have assigned affective meaning to.

\subsection{Limitations and Future Work}
Despite the promising results, several limitations should be acknowledged. First, both studies primarily sampled younger adults, and the second study was limited to students. While the resulting map is robust for this demographic, it may not generalize fully to older populations or individuals from different educational or cultural backgrounds. Future work should validate this mapping across more diverse samples. 

While the use of self-assessment might suggest a risk of subjective bias, this is a deliberate and central feature of the methodology, as the goal is to create a mapping grounded in the internally consistent, subjective experience of each individual, not to validate the animations as objective stimuli.

The creative ``palette'' offered to users was limited to a predefined set of shapes and animation types. Expanding the tool to allow for more complex creations, such as multiple interacting objects or user-drawn motion paths, could enable even more nuanced emotional expression. Furthermore, a study with a broader range of labels should be performed.

The most critical next step is finally the practical application of this map. As mentioned in the introduction, this map can be a valuable tool for harmonizing machine learning datasets. Future work should involve using the empirically derived VAD-to-label correspondence to convert annotations from different datasets. This will enable the creation of larger, more powerful corpora for training next-generation emotion recognition models.

\section{Conclusions}\label{sec:conclude}
This paper introduces and empirically validates a novel, proxy-based method for addressing the longstanding challenge of mapping between discrete and dimensional models of emotion. Using simple, user-generated geometric animations as an intermediary, we demonstrated that a reliable and psychologically grounded translation can be created between emotion labels and the VAD space. An iterative, two-study process confirmed the method's feasibility and robustness. This process showed that the core findings remained consistent even as the user interface was refined to allow for greater expressive freedom. This work's primary contribution is the resulting comprehensive emotion map. The map was shown to be stable and aligned with established affective theories. It provides a definitive set of VAD coordinates for a range of common emotions.

This research has practical and conceptual implications. The final emotion map is a valuable resource for the affective computing community because it provides an immediate tool for harmonizing disparate datasets used for training machine learning models. For example, in multimodal fusion, the map allows a discrete emotion recognized from facial expressions (e.g., using the Aff-Wild2 dataset~\cite{Kollias2019Sep}) to be converted into its VAD space, enabling direct integration with dimensional outputs from physiological signals (e.g., from the DEAP dataset~\cite{Koelstra2011Jun}).
More broadly, this work promotes a human-centric approach to emotion mapping by grounding the process in the user's creative expression and perceptual judgment cycles. By successfully using a dynamic, nonverbal artifact to bridge the gap between qualitative labels and quantitative dimensions, this method paves the way for the development of technologies that can better understand and model the nuances of human emotion.

\bibliographystyle{IEEEtran}
\bibliography{bibliography.bib}

\end{document}